\documentclass{article}
\PassOptionsToPackage{numbers,sort&compress}{natbib}
\usepackage[preprint]{neurips_2026}
\newcommand{\equalcontrib}{$^*$}
\newcommand\blfootnote[1]{%
  \begingroup
  \renewcommand\thefootnote{}\footnote{#1}%
  \addtocounter{footnote}{-1}%
  \endgroup
}

\usepackage{array}
\newcolumntype{C}[1]{>{\centering\arraybackslash}m{#1}}
\usepackage{pifont}
\usepackage{caption}
\usepackage{longtable}
\usepackage{booktabs}
\usepackage{placeins}
\usepackage[utf8]{inputenc}
\usepackage[T1]{fontenc}
\usepackage{hyperref}
\usepackage{url}
\usepackage{amsfonts}
\usepackage{nicefrac}
\usepackage{microtype}
\usepackage{xcolor}
\usepackage{graphicx,multicol,multirow,adjustbox,amsmath,amssymb,algorithm,algpseudocode}
\usepackage[most]{tcolorbox}

\title{How Well Do Large Language Models Capture Human Personality?}

\author{
\parbox{\linewidth}{\centering
Aanisha Bhattacharyya\equalcontrib\ \textsuperscript{\includegraphics[height=1.4em]{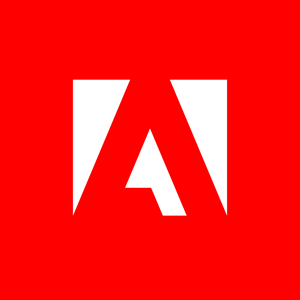}}\ \textsuperscript{\includegraphics[height=1em]{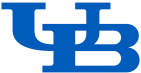}}\ \textsuperscript{\includegraphics[height=1em]{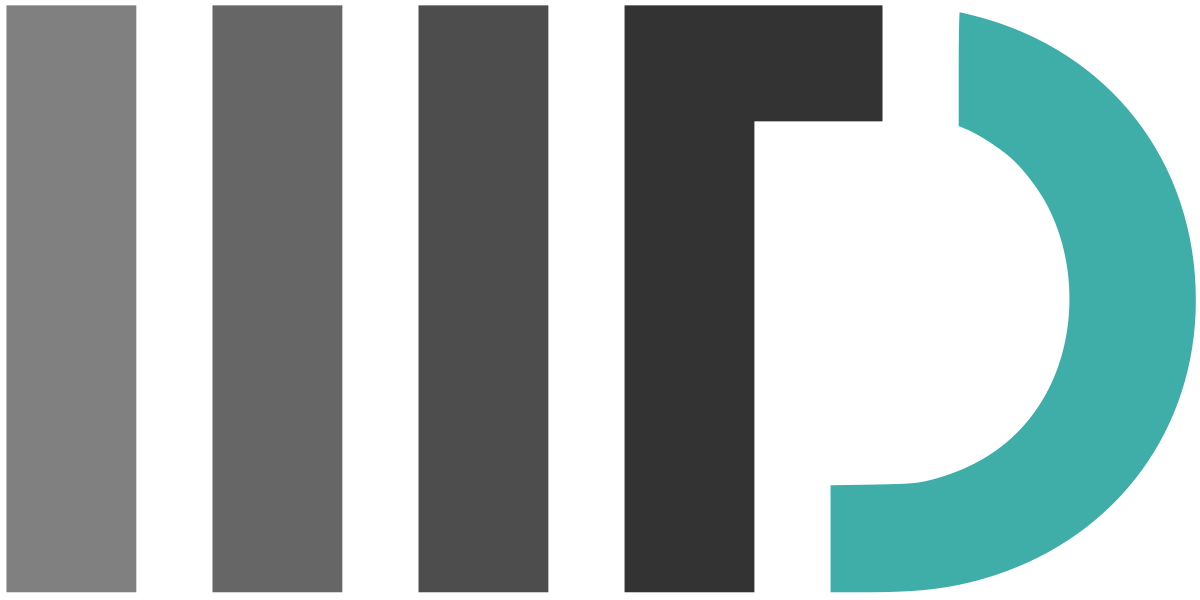}}\quad
Yaman Kumar Singla\equalcontrib\
\textsuperscript{\includegraphics[height=1.4em]{figs/adobe-logo.png}}\\
\textbf{Rajiv Ratn Shah}\
\textsuperscript{\includegraphics[height=1em]{figs/iiitd-logo.png}}\quad
\textbf{Changyou Chen}\
\textsuperscript{\includegraphics[height=1em]{figs/ub-logo.png}}\quad
\textbf{Jitendra Ajmera}\
\textsuperscript{\includegraphics[height=1.4em]{figs/adobe-logo.png}}
\\[1em]
\textsuperscript{\includegraphics[height=1.4em]{figs/adobe-logo.png}}
Adobe Media and Data Science Research (MDSR)\\
\textsuperscript{\includegraphics[height=1em]{figs/iiitd-logo.png}} IIIT-Delhi, 
\textsuperscript{\includegraphics[height=1em]{figs/ub-logo.png}}
SUNY at Buffalo\\
\texttt{\href{mailto:behavior-in-the-wild@googlegroups.com}{\textcolor{magenta}{behavior-in-the-wild@googlegroups.com}}}
}}

\begin{document}

\maketitle
\blfootnote{\equalcontrib \small Equal Contribution. Contact behavior-in-the-wild@googlegroups.com for questions and suggestions.}

\begin{abstract}

  Large language models (LLMs) are increasingly used to simulate human populations via persona prompting, often under the assumptions that richer persona descriptions improve behavioral fidelity, similarly sized attribute combinations are equally simulatable, and persona definitions generalize across tasks. In this work, we formalize these assumptions and systematically evaluate them across multiple architectures, scales, and simulation settings. We identify a fundamental limitation we term \textbf{persona manifold collapse}, where increasingly expressive persona specifications lead to systematic contraction of representational and behavioral diversity. Across models, increasing persona complexity consistently reduces inter-persona separation in latent space and weakens behavioral differentiation in downstream simulation tasks. These effects persist across multiple analyses as richer personas fail to preserve human subgroup disagreement, performance varies across attribute combinations of similar size, and adding descriptive detail often degrades rather than improves simulation fidelity. Surprisingly, simple Age--Gender personas consistently outperform richly specified Ideal Customer Profiles (ICPs) across industries, achieving substantially higher downstream prediction accuracy. We find that collapse is not uniform across attributes. Certain combinations remain behaviorally stable and preserve stronger alignment with human responses, forming localized regions we term \emph{alignment bridges}. Together, our results provide empirical and conceptual foundations for understanding the limits of persona-conditioned simulation, highlighting the need for representation-aware persona construction rather than increasing persona expressivity alone.
  
\end{abstract}

\section{Introduction}

Recent literature reflects heightened enthusiasm around persona prompting and LLM personalization, alongside a growing body of work exploring applications in automated human studies~\citep{10.1145/3708319.3733685, argyle2023out, aher2023using, horton2023large, hewitt2024predicting, park2024generative, manning2024automated, brand2024using, li2024using}, human behavior simulation~\citep{bhattacharyya2026social, kolluri2024finetuning}, personalization~\citep{10.1145/3589335.3651520, moon2024virtual}, user modeling~\citep{bougie2025simuser, ma2025pub, 10.1145/3772363.3799039}, design ideation~\citep{hamalainen2023evaluating, 10.1145/3746059.3747769}, and data generation~\citep{frohling-etal-2025-personas, nemotron_personas, ge2025scalingsyntheticdatacreation, wang2025deeppersonagenerativeenginescaling}. Across these settings, persona prompting is increasingly used to construct synthetic populations that act as proxies for real users and participants.

In this paradigm, models are conditioned on demographic personas and treated as synthetic respondents capable of generating survey answers at scale. Researchers construct ``digital twins'' of human respondents using demographic attributes and prompt LLMs to generate responses on their behalf, effectively replacing traditional survey collection with synthetic sampling \citep{10.1145/3708319.3733685, argyle2023out}. Building on this premise, LLMs have been used to recover canonical findings in behavioral economics and social psychology \citep{horton2023large, aher2023using}, predict responses on the General Social Survey and Big Five inventories \citep{park2024generative}, and simulate participant behavior in social-science experiments, treating the model as a proxy population whose aggregate behavior approximates outcomes in unseen scientific studies \citep{hewitt2024predicting}. Together, these works advance the claim that human samples used in social-science research can be substantially replaced by persona-conditioned synthetic respondents.

Beyond automating human studies, similar ideas have also been adopted for downstream applications, where persona-conditioned models serve as proxies for diverse user populations \citep{hamalainen2023evaluating}. Persona-conditioned agents have similarly been used in market research to elicit willingness-to-pay and replicate consumer experiments \citep{brand2024using, li2024using}, in recommender-system evaluation to simulate clicks, ratings, and multi-turn dialogue in place of live users \citep{bougie2025simuser, ma2025pub}, and in automated A/B testing, where structured persona agents navigate live webpages and aggregate outcomes across simulated populations to estimate treatment effects prior to deployment \citep{10.1145/3772363.3799039}. Persona conditioning has also been applied to audience-targeted content generation, where LLM-written advertisements match or surpass human-written ones in influencing user engagement \citep{10.1145/3589335.3651520}, and more broadly to social simulations in which agents emulate human behaviors, preferences, and judgments \citep{bhattacharyya2026social}. Across these settings, persona prompting has emerged as a common primitive for substituting synthetic for human input when recruiting real participants is slow, costly, or otherwise constrained.

Whereas the applications described above use persona agents to substitute for human input in downstream studies, content generation, and product evaluation, persona prompting has increasingly begun to shape the development of LLMs themselves. In particular, large-scale synthetic persona datasets such as the \textbf{Nemotron Personas} dataset~\citep{nemotron_personas} extend this paradigm to training and evaluation by constructing personas from demographic, contextual, and behavioral attributes, such as age, country, education level, career goals, hobbies, and internet usage patterns. These datasets comprise hundreds of thousands to millions of synthetic personas defined over structured attribute spaces (e.g., 22 persona and contextual traits), enabling systematic coverage of population diversity and controlled training and evaluation across diverse persona types.

Beyond structured attribute based personas, personas have been extended to expressive narrative forms encoding psychological traits, preferences, values, and lived experiences inferred from structured web knowledge, online profiles, LLM chat logs, and long-term interaction histories. PersonaHUB~\citep{ge2025scalingsyntheticdatacreation} constructs large-scale persona collections from web knowledge and uses them to generate diverse synthetic personas. \textbf{DEEPPERSONA}~\citep{wang2025deeppersonagenerativeenginescaling} further argues that existing personas are ``shallow and simplistic,'' introducing personas with hundreds of structured attributes and long-form profiles approaching 1MB of text. Building on these ideas, recent systems instantiate agents through long biographical narratives encoding decades of relationships, beliefs, motivations, and experiences. These personas are increasingly used across dialogue systems and simulations, and also for alignment, where synthetic populations replace the human respondents that traditionally provide preferences, judgments, and feedback (RLAIF). These developments mark a shift where LLM-based personas no longer merely simulate users downstream, but increasingly shape the data and feedback signals used to build LLMs themselves.

Yet as personas move from simulating users to shaping the models themselves, the assumptions underlying this paradigm become increasingly consequential. Several such assumptions, though rarely made explicit, structure how persona-based simulation is designed and interpreted.

\textbf{Expressivity} A central assumption is that increasing the descriptive richness of personas improves simulation fidelity. This motivates the construction of highly expressive personas in prior work. For instance, DEEPPERSONA~\citep{wang2025deeppersonagenerativeenginescaling} explicitly argues that existing personas are “shallow and simplistic” and introduces narrative-complete personas with rich psychological traits, preferences, and life histories to improve alignment and task performance. Similarly, Nemotron Personas~\citep{nemotron_personas} are designed around increasing attribute richness for “behavioral realism,” incorporating structured demographic fields along with rich narrative components such as \textbf{career goals}, \textbf{skills}, and \textbf{hobbies}. Persona generation approaches based on long-form social data further emphasize richer conditioning to improve emotional and behavioral realism.

\textbf{Attribute Fidelity.} Another implicit assumption is that all persona attribute combinations of the same size are similarly simulatable by LLMs. Concretely, if a model can faithfully simulate one 3-attribute persona (e.g., defined by a particular value tuple\{education, income, race\}), it is implicitly assumed that it should also simulate all other 3-attribute personas (e.g., \{income, political affiliation, race\}) with comparable fidelity. This assumption appears in large-scale persona generation works as PersonaHUB~\citep{ge2025scalingsyntheticdatacreation}, which samples personas from fixed attribute schemas and treats them as interchangeable generators across downstream tasks, and Nemotron~\citep{nemotron_personas}, which constructs large persona pools using the same 22 persona and contextual attributes for training, evaluation, and safety testing.

\textbf{Specificity.} A related assumption is that adding more attributes to a persona, improves simulation fidelity. Concretely, if a model faithfully simulates a persona defined by $N$ attributes, adding an additional attribute is expected to provide more specific behavioral grounding rather than degrade performance. This assumption motivates progressive persona enrichment in prior works like \textbf{DEEPPERSONA}~\citep{wang2025deeppersonagenerativeenginescaling} incrementally expands personas using structured taxonomies containing hundreds of attributes, while Nemotron~\citep{nemotron_personas} explicitly favors richer personas defined over 22 demographic, contextual, and behavioral fields. Other persona-generation frameworks similarly rely on multi-stage attribute expansion under the premise that greater specificity leads to more faithful simulation.

\textbf{Task Generalization.} Finally, persona definitions are often assumed to generalize across tasks. PersonaHub~\citep{ge2025scalingsyntheticdatacreation} reuses the same personas across diverse tasks such as mathematical reasoning, QA, and generation, while Nemotron~\citep{nemotron_personas} applies fixed personas across training, evaluation, and safety testing. Survey simulation works further assume that personas conditioned on demographic attributes generalize across domains such as opinion prediction, narrative generation, and behavioral tasks.

Together, literature reflects a common view that richer, larger, and more structured personas lead to more faithful and generalizable simulation of human behavior. As persona prompting is increasingly deployed in settings such as automated human studies, design ideation, and A/B testing, its outputs can influence both scientific conclusions and the content surfaced to users. Yet these assumptions are rarely systematically validated, despite entire pipelines being built on top of them. This raises a critical question: do assumptions about persona fidelity actually hold? Further, when personas appear plausible to humans, it remains unclear whether LLMs meaningfully interpret and act on them in a consistent and behaviorally faithful manner.

We conduct two complementary analyses to evaluate whether the assumptions underlying persona-based simulation hold in practice. First, we study the latent representation of personas by analyzing how persona embeddings evolve as attributes are incrementally added. This allows us to directly test assumptions about attribute enrichment and specificity, if richer personas provide more faithful behavioral grounding, persona representations should become more distinct and behaviorally separable as additional attributes are introduced. Second, we perform empirical validation through downstream simulation tasks, assessing whether persona-conditioned agents preserve human opinion differences across demographic subgroups and whether these representations translate into faithful simulation across tasks.

For the first experiment, we construct personas through hierarchical attribute composition, ranging from minimal two-attribute specifications such as Age, Gender to progressively richer profiles incorporating attributes like Education, Decision Style, and Background. We then extract persona-conditioned hidden-state embeddings across subjective and preference-oriented prompts (details in table ~\ref{tab:persona_prompt_examples}) We define \textbf{persona distance} as the mean pairwise Euclidean distance between persona embeddings at each level of attribute enrichment, using it as a proxy for behavioral separation between personas. Under the standard assumption that richer personas encode more specific and distinctive behavioral information, adding new attributes should either increase separation between persona representations or leave existing distinctions unchanged. Intuitively, if two personas already differ along attributes such as age and gender, introducing additional information such as education, decision style, or background should further refine these differences rather than collapse them into more similar representations. 

Contrary to this expectation, we observe a systematic contraction of the persona manifold as additional attributes are introduced. On Qwen-72B-Vision-Instruct, the mean persona distance drops from $14.38$ for minimal personas to $5.90$ for the richest configurations, a reduction of nearly 60\% (Table~\ref{tab:pairwise_distance_persona}). Similar declines appear across Qwen3-8B and LLaMA-3.2-90B-Vision-Instruct, indicating robustness across architectures and scales. Figure~\ref{fig:mylabel} visualizes this progressive collapse in embedding space. We term this phenomenon \textbf{persona manifold collapse}: increasing persona complexity drives representations toward narrower and more homogeneous latent regions, reducing rather than expanding behavioral diversity. 

To test whether persona manifold collapse extends beyond our controlled setup, we evaluate \textbf{Nemotron}~\citep{nemotron_personas} and \textbf{PersonaHub}~\citep{ge2025scalingsyntheticdatacreation} by measuring mean pairwise persona distances (Table~\ref{tab:persona_dataset_comparison}). Both datasets are constructed under the assumption that richer and more structured personas improve simulation fidelity, with PersonaHub relying on large-scale fixed attribute schemas and Nemotron emphasizing attribute richness for behavioral realism. Despite this, both datasets exhibit substantially lower latent separation than minimal two-attribute personas. For example, on Qwen-72B, \textbf{Nemotron} achieves a mean distance of only $5.25$, compared to $14.38$ for simple Age--Gender personas under the same model. These results indicate that increasing descriptive richness and attribute complexity do not necessarily produce more diverse or behaviorally separated representations, but instead reproduce the same collapse pattern observed in our controlled experiments. Additional ablations show that persona manifold collapse cannot be fully explained by attention saturation from long prompts or sensitivity to superficial paraphrasing. Persona representations remain relatively stable across large variations in prompt length and semantically equivalent reformulations, suggesting that the collapse arises more fundamentally from representational interference between attributes rather than prompt formatting effects (Tables~\ref{tab:attention_saturation}, and~\ref{tab:prompt_sensitivity}). Details in Sec ~\ref{exp_1}.

To further examine the behavioral consequences of \textbf{persona manifold collapse}, we test whether persona-conditioned LLMs preserve disagreement between real human subpopulations across socio-political opinion (\textbf{OpinionQA}), moral reasoning (\textbf{Moral Machine}), and aesthetic preference (\textbf{Website Likability}) tasks \citep{pmlr-v202-santurkar23a, awad2018moral, reinecke2014quantifying}. If persona-conditioned models preserved human demographic variation, subgroup pairs with strong disagreement in human annotations would also remain behaviorally separated in model outputs, resulting in strong positive correlations. Instead, correlations remain weak or negative across all tasks (Tables~\ref{tab:pearson_corr} and~\ref{tab:opinion_moral_corr}). For example, GPT-4o reaches $-0.37$ on \textbf{Website Likability}, while LLaMA-3.2-90B-Vision-Instruct reaches approximately $-0.30$ on \textbf{Moral Machine}. These results indicate that demographic groups exhibiting strong disagreement in human populations often collapse toward similar persona-conditioned outputs, limiting the ability of LLMs to preserve fine-grained behavioral diversity. Details in Sec ~\ref{exp_2}.

In summary, our results challenge several core assumptions underlying persona-conditioned simulation. Increasing persona specificity does not reliably improve behavioral fidelity, similarly sized attribute combinations are not equally simulatable, and personas that appear effective in one domain do not consistently generalize across tasks. Instead, across embedding analyses, downstream behavioral evaluations, and real-world prediction tasks, we observe consistent evidence of \textbf{persona manifold collapse}, where increasingly rich persona specifications reduce representational and behavioral diversity. However, these findings do not imply that persona prompting is uniformly ineffective. We find that the collapse is not equally severe across all attributes and attribute combinations. Certain combinations remain substantially more stable than others and preserve stronger behavioral separation and alignment with human responses. These results suggest that effective persona design depends less on maximizing persona richness and more on identifying behaviorally stable attribute combinations.

We first examine this phenomenon in a marketing and user-behavior simulation setting. Prior work on persona-conditioned LLM populations has reported competitive performance on tasks such as opinion QA, website preference evaluation, and advertising response prediction~\citep{bhattacharyya2026social}, motivating increasingly expressive persona specifications such as Ideal Customer Profiles (ICPs) that encode demographic, psychographic, and behavioral attributes. Under the standard assumption that richer personas improve behavioral fidelity, ICPs should outperform simple demographic personas. However, we observe the opposite trend. Across industries, minimal Age--Gender personas consistently outperform both auto-generated and expert-defined ICPs, achieving an average accuracy of $61.80\%$, compared to $52.66\%$ for auto-generated ICPs and $50.74\%$ for expert-defined brand ICPs (Tables~\ref{tab:em_results} and~\ref{tab:agent_industry_results}). These simple demographic personas therefore emerge as \emph{alignment bridges}, suggesting that effective simulation depends less on maximal persona expressivity and more on whether the selected attributes align with stable latent factors represented by the model. Supporting this, our ablation experiments show that personas with strong alignment remain consistently strong even after substantial elaboration, while weak personas remain weak, indicating that stable attribute configurations preserve their relative behavior despite increasing descriptive complexity (Table~\ref{tab:persona_elaboration}). Details in Sec ~\ref{sec:collapse_vs_simulation}

Taken together, our results challenge several core assumptions underlying persona-conditioned simulation. Increasing descriptive depth does not reliably improve alignment, personas with the same number of attributes do not exhibit comparable fidelity, adding additional attributes can substantially degrade behavioral separation, and personas that appear effective in one domain often fail to generalize across tasks. Across latent-space analyses and downstream behavioral evaluations, we observe a consistent pattern of \textbf{persona manifold collapse}, where increasing persona complexity contracts rather than expands effective behavioral diversity. At the same time, this collapse is not uniform across attributes: certain attribute combinations remain comparatively stable and continue to preserve stronger alignment with human behavior. These \emph{alignment bridges} suggest that effective persona design depends less on maximizing expressivity and more on identifying stable, behaviorally meaningful attribute configurations that models can reliably represent.

\vspace{-10pt}
\section{Experiments}
\label{sec: exp_main}
We conduct three complementary experiments to examine the limits of persona-conditioned simulation in large language models. First, we analyze how persona embeddings evolve as attributes are incrementally combined, probing whether richer personas expand or contract the latent representation space. Second, we evaluate whether persona-conditioned LLMs preserve human behavioral variation across demographic subgroups across diverse tasks. Third, we study how these representational effects translate to downstream simulation performance in real-world marketing and user-behavior prediction tasks. Detailed experimental protocols and results are presented in Sec.~\ref{sec:app_exp}.

\subsection{Investigating the Persona Manifold Collapse}
\label{exp_1}

\begin{figure*}[t]
    \centering
    \includegraphics[width=0.95\textwidth]{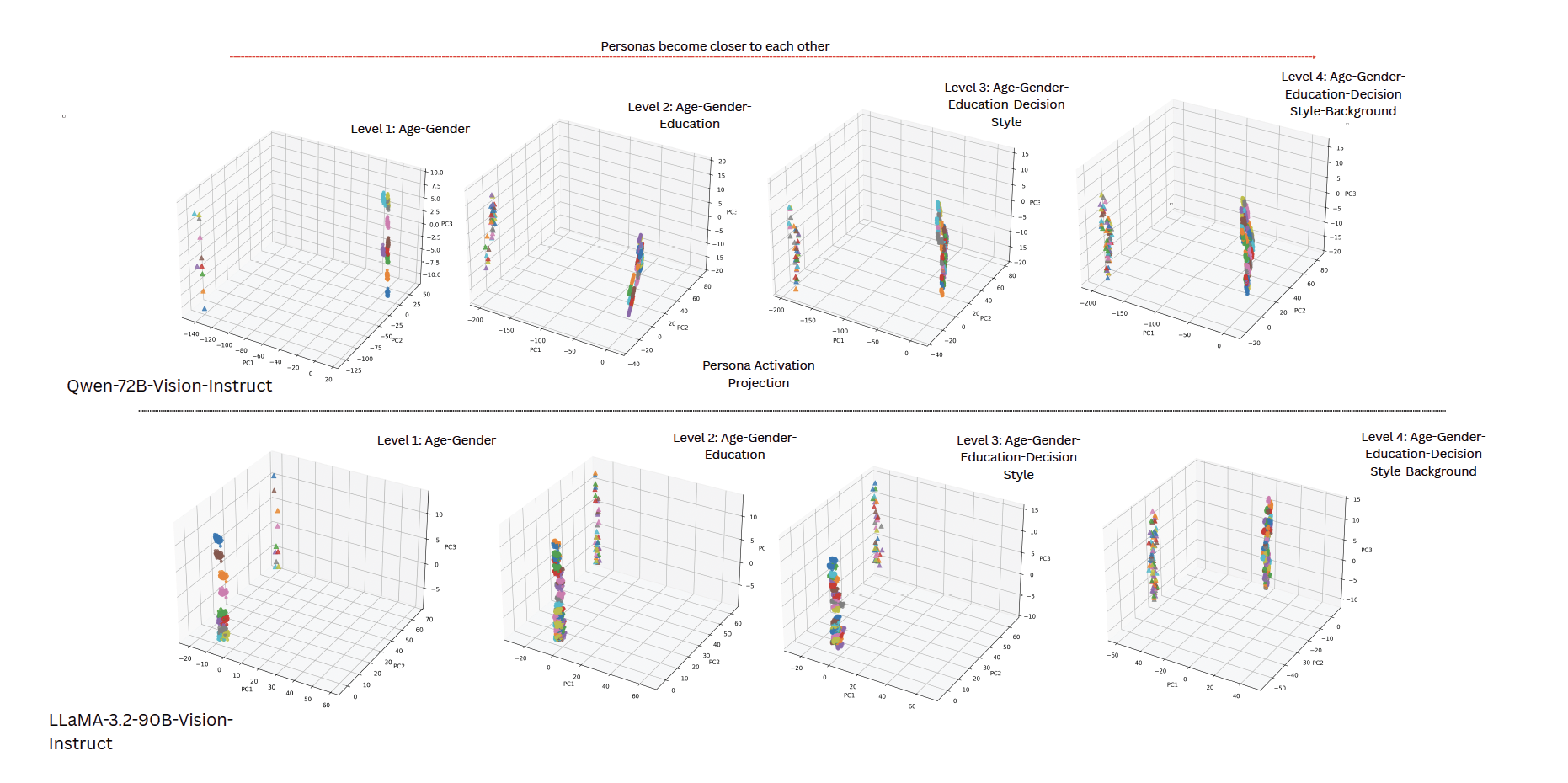}
    \caption{Persona activation vectors are projected into the first three principal components for two representative models, Qwen-72B-Vision-Instruct (top) and LLaMA-3.2-90B-Vision-Instruct (bottom), as persona specifications become progressively richer: Level~1 (Age--Gender), Level~2 (+Education), Level~3 (+Decision Style), and Level~4 (+Background). As additional attributes are introduced, persona embeddings contract toward a narrow region of latent space, indicating systematic \emph{persona manifold collapse}. Quantitatively, for Qwen-72B-Vision-Instruct, the mean pairwise distance decreases from $14.38$ at Level~1 to $9.79$ at Level~4 and further to $5.41$ at Level~5 (Table~\ref{tab:pairwise_distance_persona}). Visualizations across increasing population sizes (10, 20, 40, and 80 personas) illustrate that this contraction persists as the number of personas grows. The consistent collapse observed across both models indicates that this phenomenon is not an artifact of model architecture or scale, but reflects a fundamental limitation of persona conditioning in large language models.}
    \label{fig:mylabel}
\end{figure*}

\textbf{Experimental Setup:} To systematically investigate persona manifold collapse, we design a controlled embedding-space analysis in which persona complexity is incrementally increased by adding attributes of greater semantic richness, from minimal demographic descriptors (e.g., age and gender) to combinations of demographic and psychographic factors. This provides a principled way to probe how persona expressivity shapes latent geometry. If richer personas induce distinct internal states, the representational manifold should expand, increasing inter-persona distances. Conversely, contraction of these distances as attributes are added directly indicates persona manifold collapse. We quantify this effect by measuring pairwise inter-persona distances across successive enrichment levels.

\textbf{Persona Construction.}
We construct personas using a hierarchical additive attribute scheme inspired by marketing, audience segmentation, and social-science survey design. Personas are progressively enriched from simple demographic attributes to more detailed behavioral and psychographic descriptions, allowing controlled analysis of how increasing persona complexity affects latent representations. At each level, we add one new attribute dimension, producing progressively richer persona specifications through additive composition. We then enumerate valid combinations of attribute values to construct persona populations with increasing semantic richness. Figure~\ref{fig:mylabel} illustrates this construction process, and representative persona prompts are provided in Appendix Sec.~\ref{eg-persona-prompts}.

\textbf{Persona Representation.}
For each persona, we prompt the language model with a fixed set of subjective and preference-oriented queries spanning domains such as advertising perception, web aesthetics evaluation, lifestyle and attitudinal judgment. The detailed set of questions is given in Appendix Sec ~\ref{perp_quest}. We obtain a vector representation for each persona by extracting the final-layer hidden states corresponding to the generated responses and averaging them across all queries. This yields a single embedding vector per persona, capturing the aggregate effect of persona conditioning on the model’s internal representations.  

\textbf{Quantifying Manifold Collapse.}
Let $\ell \in \{1,\dots,L\}$ denote the persona complexity level, and let $\mathbf{V}^{(\ell)} = \{\mathbf{v}^{(\ell)}_1, \mathbf{v}^{(\ell)}_2, \dots, \mathbf{v}^{(\ell)}_{N_\ell}\}$ denote the set of persona embeddings constructed at level $\ell$, where $N_\ell$ is the number of personas at that level. For each level, the pairwise Euclidean distance matrix $\mathbf{D}^{(\ell)} \in \mathbb{R}^{N_\ell \times N_\ell}$ is computed, with entries
\begin{equation}
    D^{(\ell)}_{ij} = \|\mathbf{v}^{(\ell)}_i - \mathbf{v}^{(\ell)}_j\|_2.
\end{equation}
Representational diversity at level $\ell$ is summarized using the mean and standard deviation of the entries of $\mathbf{D}^{(\ell)}$. 

\textbf{Findings:} Under the hypothesis that increasing persona richness through additional attributes expands behavioral expressivity, these distances should increase monotonically with $\ell$, reflecting a widening of the persona manifold. Instead, we observe a systematic contraction of inter-persona distances as $\ell$ increases (Table~\ref{tab:pairwise_distance_persona}), indicating that progressively richer persona specifications lead to increasingly similar latent representations. Across models, the magnitude of this collapse is substantial, reaching $53.95\%$ for Qwen3-8B, $58.93\%$ for Qwen-72B-Vision-Instruct, and $54.70\%$ for LLaMA-3.2-90B-Vision-Instruct, while even base models exhibit consistent $20$--$30\%$ contraction. We further find that this behavior extends beyond our controlled personas to large-scale persona datasets such as \textbf{Nemotron} and \textbf{PersonaHub}, both of which exhibit substantially lower latent separation than simple Age--Gender personas despite relying on richer attribute schemas (Table~\ref{tab:persona_dataset_comparison}). Together, these results constitute direct empirical evidence of \emph{persona manifold collapse}. Additional ablations investigating attention saturation, prompt sensitivity, and attribute-level stability (Tables~\ref{tab:attention_saturation},~\ref{tab:length_accuracy},~\ref{tab:prompt_sensitivity},~\ref{tab:alignment_bridges}, and~\ref{tab:stable_unstable_personas}) show that the collapse cannot be explained solely by longer prompts or superficial prompt reformulations, and that certain attribute combinations remain substantially more stable than others across tasks and models.

\subsection{Testing Human Opinion Variance via Inter-Persona Similarity Trends}
\label{exp_2}

\textbf{Experimental Setup:} We evaluate persona-conditioned simulation across three domains spanning socio-political reasoning, moral judgment, and visual preference. \textbf{OpinionQA} evaluates alignment with human responses on socio-political issues such as taxation, immigration, and climate policy, where opinions vary systematically across demographic groups \citep{pmlr-v202-santurkar23a}. \textbf{Moral Machine} evaluates persona-conditioned decision making in trolley-style moral dilemmas with strong subgroup differences across country, education, religiosity, and political orientation \citep{awad2018moral}. \textbf{Website Likability} measures demographic variation in aesthetic preference by asking models to predict human likability scores for website screenshots \citep{reinecke2014quantifying}.

Together, these tasks provide complementary settings for evaluating whether persona-conditioned LLMs preserve behavioral variation across demographic subgroups. All tasks contain large-scale human annotations, enabling direct comparison between real population disagreement and LLM-based persona simulations. Human subgroups are defined using combinations of demographic, socioeconomic, behavioral, and geographic attributes, including age, gender, education, income, profession, political orientation, religiosity, and country (Table~\ref{tab:demographic_attributes}). Prior work on LLM-based social simulation has similarly used these tasks to evaluate persona-conditioned agents \citep{bhattacharyya2026social, pmlr-v202-santurkar23a, Bhattacharyya_Agrawal_Singla_Menta_Sr_Shah_Chen_Krishnamurthy_2026}.

\textbf{Identifying Population Disagreement Pairs:} Let $\mathcal{A} = \{a_1, a_2, \dots, a_K\}$ denote the set of available persona attributes, where each attribute $a_k$ takes values from a finite domain $\mathcal{V}_k$. A demographic subgroup $g$ is defined as a conjunction of up to four attribute-value assignments:
\begin{equation}
g = \{(a_{i_1} = v_{i_1}), \dots, (a_{i_m} = v_{i_m})\}, \quad m \leq 4,
\end{equation}
where $a_{i_j} \in \mathcal{A}$ and $v_{i_j} \in \mathcal{V}_{i_j}$. This construction yields a large collection of population subgroups spanning diverse demographic, socioeconomic, behavioral, and political characteristics.

For each subgroup $g$, we compute its empirical behavioral profile from human annotations within a given task. Depending on the task, this profile corresponds to distributions of website ratings, socio-political survey responses, or moral-decision preferences. We then measure disagreement between two subgroups $(g_p, g_q)$ by computing the distance between their empirical behavioral profiles over the set of shared evaluation instances:
\begin{equation}
\Delta(g_p, g_q) = \frac{1}{|\mathcal{S}_{pq}|} \sum_{s \in \mathcal{S}_{pq}} d\big(P_{g_p}(s), P_{g_q}(s)\big),
\end{equation}
where $\mathcal{S}_{pq}$ denotes the shared evaluation set, $P_{g}(s)$ denotes the empirical response distribution of subgroup $g$ on instance $s$, and $d$ denotes the task-specific distributional distance metric. We exhaustively enumerate valid subgroup pairs and rank them according to $\Delta(g_p, g_q)$, selecting highly divergent subgroup pairs as test cases for evaluating whether persona-conditioned LLMs preserve fine-grained human behavioral differences. Representative subgroup attributes are listed in Table~\ref{tab:demographic_attributes}.

\textbf{LLM-Based Population Simulation.}
For each high-divergence subgroup pair $(g_p, g_q)$, we convert structured attribute specifications into natural-language persona prompts and use them to condition LLM-based agents. Depending on the task, agents generate website likability ratings, socio-political survey responses, or moral decisions, enabling direct comparison with human subgroup behavior. Following prior work \citep{bhattacharyya2026social, pmlr-v202-santurkar23a}, models are evaluated using task-specific prompting protocols with in-context examples where appropriate.

\textbf{Quantifying Human--LLM Behavioral Divergence.}
Let $\mathcal{S} = \{s_1, \dots, s_N\}$ denote the set of evaluation instances and let $(g_p, g_q)$ be a selected subgroup pair. For each subgroup and evaluation instance, we obtain both human responses and persona-conditioned LLM predictions. We then compute subgroup-level behavioral disagreement by measuring distances between empirical response distributions across shared evaluation instances:
\begin{equation}
D(g_p, g_q) = \frac{1}{|\mathcal{S}|} \sum_{s \in \mathcal{S}} d\big(P_{g_p}(s), P_{g_q}(s)\big),
\end{equation}
where $P_g(s)$ denotes the subgroup response distribution on instance $s$, and $d$ denotes the task-specific distributional distance metric. For \textbf{OpinionQA}, following \citet{pmlr-v202-santurkar23a}, we additionally measure alignment between subgroup-level human responses and persona-conditioned model predictions.

\textbf{Human--LLM Distance Correlation.}
To evaluate whether persona-conditioned models preserve behavioral variation across demographic populations, we compute correlations between human subgroup disagreement and LLM subgroup separation:
\begin{equation}
\rho = \mathrm{corr}\big(\{D_H(g_p, g_q)\}, \{D_L(g_p, g_q)\}\big).
\end{equation}

\textbf{Findings:} A strong positive correlation would indicate that demographic subgroup pairs exhibiting large disagreement in human annotations also remain behaviorally separated in persona-conditioned model outputs. Instead, correlations remain consistently weak or negative across all three tasks (Tables~\ref{tab:pearson_corr} and~\ref{tab:opinion_moral_corr}). For example, GPT-4o reaches $-0.37$ on \textbf{Website Likability}, while LLaMA-3.2-90B-Vision-Instruct reaches approximately $-0.30$ on \textbf{Moral Machine}. Even the strongest positive result remains relatively weak, with Qwen3-8B-Base reaching only $0.30$ on \textbf{OpinionQA}. In practice, this means that demographic groups that humans treat as substantially different often produce very similar persona-conditioned outputs, indicating behavioral flattening across populations. The behavior also varies considerably across tasks and models, showing that persona definitions that appear effective in one domain do not reliably generalize to others, and that attribute combinations with the same number of attributes can exhibit substantially different behavioral fidelity. Together, these results provide downstream behavioral evidence of \textbf{persona manifold collapse}.

\vspace{-10pt}
\subsection{Quantifying impact of Persona Manifold Collapse on Simulation Ability}
\label{sec:collapse_vs_simulation}

We study the impact of persona complexity on downstream simulation performance in two realistic behavioral prediction tasks: (i) user engagement with brand-authored social media posts, and (ii) click-through rate prediction for marketing emails. Together, these tasks evaluate whether richer, expert-defined personas improve simulation accuracy or whether increasing persona complexity instead degrades performance despite more detailed persona specifications.

\textbf{Experimental Setup.} \textbf{Tweet Engagement Prediction.} We study tweet engagement prediction across the \emph{technology}, \emph{airlines}, and \emph{fashion} industries using brand-authored social media posts from the CBC dataset \citep{khandelwal2024large}, where the goal is to predict human engagement percentiles. \textbf{Email CTR Prediction.} We additionally evaluate click-through rate (CTR) prediction on a large-scale marketing email dataset from an industry collaboration in the creative sector, where the task is to predict whether an email achieves high or low engagement based on its content and target audience.

\textbf{Persona Construction.} For each brand (tweet) and campaign (email), we construct persona agents using Ideal Customer Profiles (ICPs) generated by GPT-5.2 with web-search augmentation. These ICPs describe target customer archetypes grounded in publicly available world knowledge and are manually verified for plausibility. We then instantiate a population of agents using detailed narrative persona prompts derived from these ICPs, enabling controlled evaluation of how rich, expert-defined persona specifications affect simulation performance. More details in Appendix Sec ~\ref{app:persona_prompts}.

\textbf{Persona-Based Simulation.}
Given a brand or campaign and its associated persona agents, each agent independently predicts outputs for all test examples. Predictions are generated by conditioning the LLM on both the task prompt and the persona narrative. For each example, we aggregate predictions across agents via simple averaging, forming an ensemble-based simulation of collective human response similar to protocol defined in \cite{bhattacharyya2026social}.

\textbf{Evaluation Metrics.}
Following the evaluation framework introduced in \citet{khandelwal2024large}, we measure simulation performance using a two-way classification accuracy metric based on a fixed engagement threshold. Specifically, both ground-truth engagement scores and aggregated persona predictions are binarized into \emph{high} and \emph{low} engagement classes using the same threshold, and accuracy is computed between predicted and true labels. We report per-industry accuracy for tweet engagement prediction and overall accuracy for the email CTR task, enabling systematic comparison across industries, domains, and persona configurations.

\textbf{Findings:} Across both tweet engagement and email CTR prediction tasks, simple demographic personas consistently outperform richer customer-profile-based personas (Tables~\ref{tab:em_results} and~\ref{tab:agent_industry_results}). In the email CTR task, Age--Gender personas achieve $70.00\%$ accuracy compared to $58.57\%$ for auto-generated ICP agents and $50.00\%$ for standard prompting baselines. Similar trends appear across all tweet engagement domains. Additional ablations show that this behavior cannot be explained solely by prompt length or attention saturation. Expanding persona descriptions to narratives exceeding 2000 tokens does not consistently improve performance, and elaborating personas preserves their relative alignment with human moral reasoning (Tables ~\ref{tab:persona_elaboration} and ~\ref{tab:length_accuracy}). Together, these results challenge the assumptions that richer personas necessarily improve behavioral fidelity or that adding more attributes and descriptive detail yields finer-grained simulation behavior, further supporting the broader pattern of \textbf{persona manifold collapse}.

\vspace{-10pt}
\subsection{Discovering Alignment Bridges in the Persona Manifold}
\label{exp_4}

\textbf{Method.} To further investigate which persona configurations remain behaviorally stable despite the broader pattern of collapse, we perform a greedy search over attribute combinations by repeatedly evaluating persona-conditioned simulations across multiple tasks and settings. Starting from simple demographic personas, we systematically vary and expand attribute compositions while tracking downstream performance and behavioral separation. This allows us to identify stable attribute combinations that consistently preserve stronger alignment with human behavior, as well as unstable combinations that repeatedly induce representational collapse and behavioral homogenization.

\textbf{Findings:} Persona manifold collapse is not uniform across attributes (Tables~\ref{tab:alignment_bridges} and~\ref{tab:stable_unstable_personas}). Certain attributes, such as education in \textbf{OpinionQA} and gender in \textbf{Moral Machine}, consistently remain more stable across combinations and act as \emph{alignment bridges}. For example, combinations such as Education + Gender and Gender + Religious preserve stronger alignment with human behavior across models. In contrast, political identity and income frequently act as collapse triggers, substantially reducing behavioral fidelity when combined with other attributes. Notably, some attributes that perform well individually degrade sharply in larger combinations, indicating that persona attributes do not combine independently or additively. Personas constructed from stable attribute combinations also exhibit substantially larger inter-persona distances than collapse-prone personas, reaching $15.78$ versus $5.88$ on Qwen-72B-VL and $7.41$ versus $2.38$ on Qwen-8B. Together, these results show that effective persona design depends not only on which attributes are used, but also on how those attributes interact within the model representation space.

\vspace{-5pt}
\section{Results and Discussion}
\label{sec:result}
\vspace{-5pt}

Overall, our experiments reveal a consistent pattern of \textbf{persona manifold collapse} across latent representations, downstream behavioral simulation, and real-world prediction tasks. Collectively, these results challenge the core assumptions that richer personas necessarily improve behavioral fidelity, that attribute combinations of similar complexity are equally simulatable, and that persona definitions generalize reliably across tasks and domains. Instead, richer persona specifications systematically reduce inter-persona separation, fail to preserve human subgroup disagreement, and often underperform simpler demographic personas in downstream tasks. At the same time, the collapse is not uniform across attributes, with certain combinations remaining substantially more stable than others. We summarize several broader implications of these findings below and refer readers to the corresponding experimental sections for detailed analysis and quantitative results.

\textbf{Persona manifold collapse is model-agnostic.} Across reasoning and non-reasoning models, spanning both LLMs and VLMs, increasing persona complexity consistently reduces behavioral diversity (Table~\ref{tab:pairwise_distance_persona}). The magnitude of collapse ranges from $22.90\%$ in Qwen3-8B-Base to $58.93\%$ in Qwen-72B-Vision-Instruct, with similarly strong contraction observed across all evaluated architectures. This consistency suggests that \textbf{persona manifold collapse} is not tied to a specific model type, scale, or architecture.

To better understand the mechanisms underlying \textbf{persona manifold collapse}, we investigate three possible causes suggested by prior work alignment-induced homogenization, sensitivity to prompt wording, and attention saturation from increasingly long persona descriptions.

\textbf{Alignment amplifies persona manifold collapse.} Prior work such as ALPHA \citep{Bhattacharyya_Agrawal_Singla_Menta_Sr_Shah_Chen_Krishnamurthy_2026} and \citet{pmlr-v202-santurkar23a} suggests that alignment can homogenize model behavior. Consistent with this, instruction-tuned models exhibit substantially stronger collapse than their base counterparts (Table~\ref{tab:pairwise_distance_persona}), with collapse increasing from roughly $35\%$ to $59\%$ in Qwen-72B and from $29\%$ to $55\%$ in LLaMA-3.2-90B.

\textbf{Effect of prompt sensitivity on persona manifold collapse.} To test whether collapse is driven by superficial prompt wording, we evaluate semantically equivalent paraphrases of the same personas while keeping all underlying attributes fixed. Pairwise persona distances remain highly stable across paraphrases for both Qwen-8B and Qwen-72B (Table~\ref{tab:prompt_sensitivity}), indicating that persona manifold collapse cannot be explained solely by prompt sensitivity. These results suggest that the specific attributes used are substantially more important than surface-level phrasing or stylistic expressivity.

\textbf{Effect of attention saturation on persona manifold collapse.} We further test whether collapse arises from increasingly long persona prompts by varying persona length from short tabular descriptions to narratives exceeding 2000 tokens while keeping attributes fixed. If attention saturation were the primary cause, performance and persona separation would degrade monotonically with length. Instead, both remain non-monotonic across prompt lengths (Tables~\ref{tab:attention_saturation} and~\ref{tab:length_accuracy}), suggesting that collapse cannot be fully explained by long-context degradation alone. Together, these results indicate that attribute composition plays a substantially more important role than prompt length or narrative detail.

\textbf{Failure is consistent across task types.}
This pattern persists across fundamentally different settings, including scalar judgments, opinion distributions, and discrete moral decisions. Despite substantial differences in output space and task structure, correlations remain uniformly weak, indicating that personas that appear behaviorally faithful in one domain do not reliably preserve demographic variation across other tasks.

\section{Conclusion}

Our results show that increasing persona richness does not reliably improve behavioral fidelity and instead often leads to \textbf{persona manifold collapse}, where representational and behavioral diversity contract as additional attributes are introduced. This pattern persists across models, tasks, and simulation settings, challenging the assumptions that richer personas necessarily yield better simulation fidelity, that similarly sized attribute combinations are equally simulatable, and that persona definitions generalize reliably across domains. However, the collapse is not uniform across attributes. Certain combinations remain substantially more stable and preserve stronger alignment with human behavior, suggesting that effective persona design depends less on maximizing persona expressivity and more on identifying behaviorally robust attribute combinations. We hope these findings motivate more representation-aware approaches to persona construction and evaluation in future work.

\appendix
\onecolumn
\section{Appendix}

\subsection{Experimental Results}
\label{sec:app_exp}
Across three complementary experimental settings, we consistently observe strong evidence of \textbf{persona manifold collapse} and its downstream consequences. First, embedding-space analyses reveal a systematic contraction of inter-persona distances as additional attributes are introduced, with richer persona specifications producing progressively more compressed latent representations rather than greater behavioral separation (Table~\ref{tab:pairwise_distance_persona}). Second, in controlled human--LLM comparison experiments spanning socio-political opinion, moral reasoning, and website likability judgments, correlations between human subgroup disagreement and persona-conditioned model divergence remain weak or negative, indicating that persona-conditioned agents fail to preserve fine-grained demographic opinion variance (Tables~\ref{tab:pearson_corr} and~\ref{tab:opinion_moral_corr}). Third, in downstream simulation tasks including advertising engagement prediction, email click-through-rate estimation, and social media response modeling, simple low-dimensional personas consistently outperform richer expert-defined persona constructions, despite using substantially less descriptive information (Tables~\ref{tab:em_results} and~\ref{tab:agent_industry_results}). Additional analyses further show that these effects cannot be explained purely by prompt length, attention saturation, or superficial prompt phrasing, but instead arise from representational interference between persona attributes (Tables~\ref{tab:attention_saturation},~\ref{tab:length_accuracy}, and~\ref{tab:prompt_sensitivity}). Together, these results establish persona manifold collapse as a robust, model-agnostic phenomenon that directly degrades behavioral fidelity in downstream simulation settings.

\subsection{Persona Simulation Questions}
\label{perp_quest}
We curate a diverse set of questions designed to elicit rich and varied responses across multiple behavioral, perceptual, and personal dimensions. These questions span sufficient breadth and depth to induce differentiated latent embeddings when answered by individuals with diverse demographic and psychographic profiles. We use this question set to evaluate whether large language models are able to preserve and reflect such diversity in their generated responses. Questions added in Table ~\ref{tab:persona_questions_1} and ~\ref{tab:persona_questions_2}.

\subsection{Example Persona Prompt Construction}
\label{eg-persona-prompts}

To illustrate how persona specifications are composed from natural-language attribute descriptions, we present representative examples of combined persona prompts generated under different attribute configurations. Each persona prompt is formed by concatenating attribute-specific narratives drawn from demographic, behavioral, and psychographic taxonomies, resulting in progressively richer and more expressive persona descriptions. These examples demonstrate the compositional structure of our persona construction pipeline and provide transparency into the semantic content used to condition the model. Table~\ref{tab:persona_prompt_examples} reports representative prompt instantiations spanning diverse attribute combinations.

\subsection{Persona Prompts and Ideal Customer Profiles}
\label{app:persona_prompts}

Table~\ref{tab:persona_longtable} presents the full set of brand-specific ideal customer profiles (ICPs) and corresponding persona prompts used in our experiments, spanning multiple commercial domains including technology, aviation, and fashion. 
For each brand, we include two representative prompts that capture distinct but realistic user archetypes relevant to the brand’s product portfolio. 
These personas are designed to reflect practical decision-making contexts, domain-specific constraints, and real-world motivations, enabling controlled evaluation of model behavior across heterogeneous consumer and enterprise settings.

\subsection{Limitations}
\label{sec:limitations}

Our study focuses primarily on persona-conditioned simulation in contemporary open and closed LLMs and VLMs, and therefore does not exhaustively cover all possible architectures, prompting strategies, or alignment procedures. While we evaluate multiple tasks spanning socio-political opinion, moral reasoning, aesthetic preference, and engagement prediction, there may exist domains where richer personas provide stronger benefits than those observed here. Finally, although we identify stable attribute combinations that partially resist collapse, we do not provide an exhaustive list of alignment bridges and collapse triggers, so there can exist better bridges, which would need more systematic exploration.

\subsection{Broader Impact}
\label{sec:broader_impact}

Persona-conditioned LLMs are increasingly used for applications such as population simulation, survey modeling, marketing analysis, and decision support. Our findings suggest that richer persona specifications do not necessarily improve behavioral fidelity and can instead reduce representational and behavioral diversity through \textbf{persona manifold collapse}. This has important implications for the reliability of simulated populations in high-stakes settings, where assumptions about demographic realism may lead to misleading conclusions or biased decisions. At the same time, our work identifies promising directions for more reliable persona construction through behaviorally stable attribute combinations. We hope these findings encourage more careful evaluation of persona-conditioned systems and motivate future work on representation-aware simulation methods.

\subsection{Experimental Compute Resources}
\label{sec:compute}

All experiments were conducted using a cluster of 8 NVIDIA A100 GPUs. A standard evaluation run, including persona generation, latent representation extraction, and downstream simulation, requires approximately 30 minutes of GPU compute time depending on the model and task setting. We use GPT-5.2 with web-search augmentation for generating ICP-based personas and GPT-4o for selected downstream evaluation and comparison experiments.

\begin{table*}[t]
\centering
\small
\setlength{\tabcolsep}{6pt}

\begin{tabular}{llcc}
\toprule
\textbf{Model} & \textbf{Persona Attribute Composition} & \textbf{Mean} & \textbf{Std.} \\
\midrule

\multirow{4}{*}{Qwen3-8B-Base}
& Age--Gender & 4.1049 & 2.6814 \\
& Age--Gender--Education & 3.8419 & 2.0556 \\
& Age--Gender--Education--Decision Style & 3.2433 & 1.8979 \\
& Age--Gender--Education--Decision Style--Background & 3.1651 & 1.4104 \\

\midrule

\multirow{4}{*}{Qwen3-8B}
& Age--Gender & 7.8125 & 1.7960 \\
& Age--Gender--Education & 4.3983 & 2.4780 \\
& Age--Gender--Education--Decision Style & 3.5126 & 1.8817 \\
& Age--Gender--Education--Decision Style--Background & 3.5976 & 1.5944 \\

\midrule

\multirow{4}{*}{Qwen-72B-Base}
& Age--Gender & 9.1269 & 6.4293 \\
& Age--Gender--Education & 7.1331 & 4.1803 \\
& Age--Gender--Education--Decision Style & 6.8915 & 4.3034 \\
& Age--Gender--Education--Decision Style--Background & 5.9719 & 4.1135 \\

\midrule

\multirow{4}{*}{Qwen-72B-Vision-Instruct}
& Age--Gender & 14.3750 & 3.8437 \\
& Age--Gender--Education & 7.5516 & 4.6803 \\
& Age--Gender--Education--Decision Style & 6.3337 & 3.7011 \\
& Age--Gender--Education--Decision Style--Background & 5.9044 & 2.4421 \\

\midrule

\multirow{4}{*}{LLaMA-3.2-90B-Vision-Base}
& Age--Gender & 8.2301 & 5.0318 \\
& Age--Gender--Education & 7.1083 & 4.1847 \\
& Age--Gender--Education--Decision Style & 6.7774 & 5.2504 \\
& Age--Gender--Education--Decision Style--Background & 5.8248 & 3.4800 \\

\midrule

\multirow{4}{*}{LLaMA-3.2-90B-Vision-Instruct}
& Age--Gender & 14.1875 & 3.8590 \\
& Age--Gender--Education & 7.8917 & 4.9873 \\
& Age--Gender--Education--Decision Style & 7.1684 & 4.1993 \\
& Age--Gender--Education--Decision Style--Background & 6.4267 & 4.0643 \\

\bottomrule
\end{tabular}

\caption{Mean and standard deviation of pairwise distances between persona embeddings as personas are progressively enriched through hierarchical attribute composition, ranging from minimal Age, Gender specifications to richer profiles incorporating Education, Decision Style, and Background. Pairwise distance is computed in the latent embedding space and serves as a proxy for behavioral separation between personas. Under the standard assumption that adding attributes increases persona specificity and behavioral distinctiveness, distances would be expected to increase or remain stable as personas become more expressive. Instead, across nearly all architectures and scales, we observe systematic \textbf{persona manifold collapse}, where increasing persona complexity contracts representations toward narrower and more homogeneous latent regions. The trend is particularly pronounced in instruction-tuned models, suggesting that richer persona specifications reduce, rather than expand, effective behavioral separation.}
\label{tab:pairwise_distance_persona}

\end{table*}

\begin{table}[t]
\centering
\small
\setlength{\tabcolsep}{8pt}
\begin{tabular}{lccc}
\toprule
\textbf{Model} & \textbf{Nemotron} & \textbf{PersonaHub} & \textbf{Age-Gender} \\
\midrule
Qwen-8B & 1.7956 & 3.8406 & 7.8125 \\
Qwen-72B & 5.2495 & 6.8355 & 14.3750 \\
\bottomrule
\end{tabular}
\caption{Mean pairwise distances between persona embeddings for widely used persona corpora and simple attribute-based personas. Despite being constructed with substantially richer and more expressive persona descriptions, curated datasets such as \textbf{Nemotron} and \textbf{PersonaHub} exhibit significantly lower separation in latent space compared to minimal Age--Gender personas. This suggests that increasing descriptive richness and attribute complexity do not necessarily produce more behaviorally distinct representations, providing further evidence of \textbf{persona manifold collapse}.}
\label{tab:persona_dataset_comparison}
\end{table}

\FloatBarrier
\clearpage
\begin{table}[t]
\centering
\small
\setlength{\tabcolsep}{4pt}

\begin{tabular}{lcccc}
\toprule
\textbf{Model} & \textbf{Persona Format} & \textbf{Tokens} & \textbf{Mean} & \textbf{Std.} \\
\midrule

\multirow{6}{*}{Qwen-72B-Vision-Instruct}
& Tabular (JSON) & 12 & 9.9364 & 5.6761 \\
& Shortest & 15 & 14.5678 & 7.5822 \\
& Short & 150 & 14.7993 & 8.4612 \\
& Paper Version & 1080 & 13.0967 & 5.6680 \\
& Medium & 1570 & 14.1680 & 7.9029 \\
& Long & 2050 & 15.3884 & 10.5479 \\

\midrule

\multirow{6}{*}{Qwen-8B}
& Tabular (JSON) & 12 & 5.8692 & 2.9637 \\
& Shortest & 15 & 7.0946 & 3.8733 \\
& Short & 150 & 7.8463 & 3.8253 \\
& Paper Version & 1080 & 5.6255 & 3.4585 \\
& Medium & 1570 & 8.9235 & 3.4073 \\
& Long & 2050 & 8.2392 & 5.3727 \\

\bottomrule
\end{tabular}

\caption{Effect of persona length and formatting on pairwise persona separation. Persona descriptions vary from compact tabular formats (12 tokens) to long-form narrative personas exceeding 2000 tokens while preserving the same core attributes. If \textbf{persona manifold collapse} were primarily caused by attention saturation or context dilution, increasing prompt length would be expected to systematically contract persona representations. Instead, pairwise distances remain relatively stable across large variations in persona length, and in several cases even increase for longer personas. These results indicate that persona manifold collapse is not merely an artifact of long prompts or limited attention capacity, but instead arises from how additional attributes interact within the latent representation space.}
\label{tab:attention_saturation}
\end{table}

\begin{table}[t]
\centering
\small
\setlength{\tabcolsep}{6pt}

\begin{tabular}{lccc}
\toprule
\textbf{Model} & \textbf{Prompt Variant} & \textbf{Mean} & \textbf{Std.} \\
\midrule

\multirow{3}{*}{Qwen-8B}
& Variant 1 & 8.5006 & 3.8609 \\
& Variant 2 & 8.2976 & 3.4696 \\
& Variant 3 & 8.4951 & 3.8699 \\

\midrule

\multirow{3}{*}{Qwen-72B}
& Variant 1 & 16.3381 & 8.6070 \\
& Variant 2 & 16.3269 & 6.3472 \\
& Variant 3 & 16.8079 & 10.9789 \\

\bottomrule
\end{tabular}

\caption{Prompt sensitivity analysis across semantically equivalent persona paraphrases. Each variant preserves the same underlying persona attributes while altering only surface-level phrasing. Mean pairwise distances remain highly consistent across paraphrases for both Qwen-8B and Qwen-72B, indicating that latent persona separation is relatively stable to minor prompt reformulations. Unlike \textbf{persona manifold collapse}, which emerges from increasing attribute complexity and representational interaction, superficial paraphrasing alone does not substantially alter the geometry of persona representations.}
\label{tab:prompt_sensitivity}
\end{table}

\FloatBarrier
\clearpage
\begin{table}[t]
\centering
\small
\setlength{\tabcolsep}{8pt}
\begin{tabular}{lc}
\toprule
\textbf{Model} & \textbf{Correlation} \\
\midrule
GPT-4o & $-0.3686$ \\
Qwen-72B-VL\_Instruct & $-0.0031$ \\
LLaMA-3.2-90B-Vision-Instruct & $0.1001$ \\
Qwen-7B-VL\_Instruct & $-0.1286$ \\
\bottomrule
\end{tabular}
\caption{Pearson correlation between human subgroup disagreement and persona-conditioned model-output separation on the \textbf{Website Likability} task. If persona-conditioned LLMs faithfully preserved human opinion variance, subgroup pairs exhibiting strong disagreement in human annotations would also remain well separated in model outputs, yielding strong positive correlations. Instead, all evaluated models exhibit weak or negative correlations, indicating substantial behavioral flattening across demographic groups. This suggests that even when human populations strongly diverge in visual preference judgments, persona-conditioned LLMs often collapse toward similar response distributions, providing downstream behavioral evidence of \textbf{persona manifold collapse}.}
\label{tab:pearson_corr}
\end{table}

\begin{table}[t]
\centering
\small
\setlength{\tabcolsep}{6pt}

\begin{tabular}{lcc}
\toprule
\textbf{Model} & \textbf{OpinionQA} & \textbf{Moral Machine} \\
\midrule
Qwen3-8B-Base & 0.2979 & 0.0887 \\
Qwen3-8B & 0.2616 & -0.0748 \\
Qwen-72B-Base & 0.1443 & -0.0661 \\
Qwen-72B-Vision-Instruct & -0.1172 & -0.2926 \\
LLaMA-3.2-90B-Vision-Base & -0.2646 & -0.2866 \\
LLaMA-3.2-90B-Vision-Instruct & -0.2646 & -0.2987 \\
\bottomrule
\end{tabular}

\caption{Pearson correlation between human subgroup disagreement and persona-conditioned model-output separation across \textbf{OpinionQA} and \textbf{Moral Machine}. \textbf{OpinionQA} evaluates socio-political opinions on issues such as taxation, immigration, and climate policy, while \textbf{Moral Machine} evaluates moral reasoning in trolley-style ethical dilemmas. Strong positive correlations would indicate that subgroup pairs exhibiting larger disagreement in human annotations also remain behaviorally distinct in model outputs. Instead, correlations remain weak or negative across most models and tasks, suggesting that persona-conditioned LLMs fail to preserve the relative structure of human disagreement. This provides downstream behavioral evidence that \textbf{persona manifold collapse} extends beyond latent representations into population-level simulation behavior.}
\label{tab:opinion_moral_corr}
\end{table}

\FloatBarrier
\clearpage
\begin{table}[t]
\centering
\small
\setlength{\tabcolsep}{6pt}

\begin{tabular}{lcc}
\toprule
\textbf{Agent Setup} & \textbf{Persona Attributes} & \textbf{Accuracy} \\
\midrule
Baseline GPT & -- & 50.00 \\
5-shot Prompting & GPT-4o Examples & 52.57 \\
Customer Agents & Auto ICP & 58.57 \\
Social Agents & Age \& Gender & \textbf{70.00} \\
\bottomrule
\end{tabular}

\caption{Performance comparison across agent configurations on the email CTR prediction task. Persona-conditioned social agents constructed using simple demographic attributes (\textbf{Age} and \textbf{Gender}) substantially outperform baseline prompting approaches and customer-agent setups, suggesting that even minimal structured personas can provide strong behavioral grounding for downstream prediction tasks.}
\label{tab:em_results}
\end{table}

\begin{table}[t]
\centering
\small
\setlength{\tabcolsep}{4pt}

\begin{tabular}{llccc}
\toprule
\textbf{Agent Setup} & \textbf{Persona Attributes} & \textbf{Fashion} & \textbf{Airlines} & \textbf{Tech} \\
\midrule
Baseline GPT & -- & 48.88 & 51.20 & 49.56 \\
Social Agents & Age \& Gender & \textbf{61.24} & \textbf{62.20} & \textbf{61.95} \\
Customer Agents & Auto ICP & 50.55 & 53.98 & 53.46 \\
Customer Agents & Brand ICP & 51.49 & 47.28 & 53.46 \\
\bottomrule
\end{tabular}

\caption{Performance across persona-based agent configurations and industry domains for the tweet engagement prediction task. Simple demographic personas based on \textit{Age} and \textit{Gender} consistently outperform both baseline prompting and customer-profile-based agent configurations across Fashion, Airlines, and Tech domains, suggesting that lightweight structured personas can provide stronger behavioral grounding than more complex customer profiling pipelines.}
\label{tab:agent_industry_results}
\end{table}

\FloatBarrier
\clearpage
\begin{table}[t]
\centering
\small
\setlength{\tabcolsep}{5pt}

\begin{tabular}{lccc}
\toprule
\textbf{Persona} & \textbf{Original} & \textbf{Elaborate} & \textbf{$\Delta$ \%} \\
\midrule
top\_1 & 0.82 & 0.78 & -4.88\% \\
top\_2 & 0.81 & 0.79 & -2.47\% \\
top\_3 & 0.80 & 0.79 & -1.25\% \\
\midrule
mid\_1 & 0.77 & 0.75 & -2.60\% \\
mid\_2 & 0.77 & 0.75 & -1.30\% \\
mid\_3 & 0.70 & 0.72 & +2.86\% \\
mid\_4 & 0.69 & 0.69 & 0.00\% \\
\midrule
bottom\_1 & 0.46 & 0.50 & +6.52\% \\
bottom\_2 & 0.46 & 0.49 & +6.52\% \\
bottom\_3 & 0.43 & 0.43 & +2.33\% \\
\bottomrule
\end{tabular}

\caption{Effect of persona elaboration on alignment with human moral reasoning. Each persona is expanded by approximately 1200--1500 additional tokens while preserving the same underlying attributes. If richer narrative detail improved representational fidelity, one would expect substantial alignment gains after elaboration. Instead, alignment changes remain marginal, and the relative ordering of personas is preserved: personas that are highly aligned with human responses remain highly aligned, while poorly aligned personas remain poorly aligned. These results suggest that increasing descriptive richness alone does not fundamentally alter behavioral fidelity, further indicating that \textbf{persona manifold collapse} is not simply resolved through longer or more elaborate persona descriptions.}
\label{tab:persona_elaboration}
\end{table}

\begin{table}[t]
\centering
\small
\setlength{\tabcolsep}{6pt}

\begin{tabular}{lcc}
\toprule
\textbf{Persona Type} & \textbf{Tokens} & \textbf{Accuracy} \\
\midrule
persona\_tabular & 12 & 58.8 \\
persona\_shortest & 15 & 63.7 \\
persona\_short & 150 & 60.0 \\
persona\_paper\_version & 1080 & 53.8 \\
persona\_medium & 1570 & 63.7 \\
persona\_long & 2050 & 58.8 \\
\bottomrule
\end{tabular}

\caption{Effect of persona length on downstream tweet engagement prediction accuracy while keeping the underlying persona attributes fixed. Persona descriptions range from compact tabular forms (12 tokens) to long-form narratives exceeding 2000 tokens. If \textbf{persona manifold collapse} primarily arose from attention saturation or context dilution, performance would be expected to degrade monotonically as persona length increases. Instead, accuracy remains non-monotonic across prompt lengths, with both extremely short (15-token) and substantially longer (1570-token) personas achieving identical peak performance. These results suggest that attention saturation alone does not explain persona manifold collapse, pointing instead toward representational interference arising from attribute composition.}
\label{tab:length_accuracy}
\end{table}

\begin{table}[H]
\centering
\small
\setlength{\tabcolsep}{5pt}

\begin{tabular}{lccc}
\toprule
\textbf{Model} & \textbf{Anchor} & \textbf{Alignment Bridge (Stable)} & \textbf{Collapse Trigger (Unstable)} \\
\midrule
Qwen-8B & Gender & Gender + Religious & Political + Income \\
Qwen-72B-VL & Gender & Gender + Political & Political + Religious \\
Qwen3-8B & Education & Education + Gender & Political + Income \\
Qwen-72B-VL & Education & Education + Race & Political + Race + Income \\
\bottomrule
\end{tabular}

\caption{Model- and task-dependent alignment bridges underlying persona manifold collapse. Alignment bridges correspond to attribute combinations that remain behaviorally stable and preserve stronger alignment with human annotations, while collapse triggers correspond to combinations that consistently destabilize persona fidelity. Across both \textbf{OpinionQA} and \textbf{Moral Machine}, stable attributes differ across models and tasks: education acts as a strong anchor in socio-political opinion modeling, whereas gender emerges as a stronger bridge in moral reasoning. In contrast, political identity and income frequently induce collapse and reduce alignment. These results suggest that persona manifold collapse is heterogeneous across attribute dimensions rather than uniformly distributed, indicating that certain attribute combinations remain representationally robust even as others collapse.}
\label{tab:alignment_bridges}
\end{table}

\begin{table}[H]
\centering
\small
\setlength{\tabcolsep}{8pt}

\begin{tabular}{lcc}
\toprule
\textbf{Model} & \textbf{Stable Personas} & \textbf{Unstable Personas} \\
\midrule
Qwen-72B-VL & 15.78 & 5.88 \\
Qwen-8B & 7.41 & 2.38 \\
\bottomrule
\end{tabular}

\caption{Inter-persona distances for personas constructed from alignment bridges (stable attribute combinations) versus collapse-triggering attributes (unstable combinations). For each model, we construct two groups of personas by selecting 10 personas composed of stable attribute combinations and 10 composed of unstable combinations, then measure mean pairwise distances within each group. Personas built from alignment bridges exhibit substantially larger separation in latent space compared to collapse-prone personas across both Qwen-72B-VL and Qwen-8B. These results indicate that \textbf{persona manifold collapse} is not uniform across attributes: certain dimensions, such as education in \textbf{OpinionQA} and gender in \textbf{Moral Machine}, remain representationally robust and act as behavioral anchors, while others such as political identity and income induce collapse and homogenization.}
\label{tab:stable_unstable_personas}
\end{table}

\begin{table}[H]
\centering
\small
\setlength{\tabcolsep}{5pt}

\begin{tabular}{l c p{10cm}}
\toprule
\textbf{Attribute} & \textbf{\# Values} & \textbf{Representative Values} \\
\midrule

Gender & 2 & \{Male, Female\} \\

Age Group & 8 & \{18--24, 25--34, 35--44, 45--54, 55+, 18--29, 30--49, 65+\} \\

Education & 12+ & \{Pre-high school, High school, College, University degree, Graduate school, Professional school, PhD, Postdoctoral, underHigh, vocational, bachelor, graduate\} \\

Income & 11+ & \{<$30k, middle, >$100k, 5k, 10k, 15k, 25k, 35k, 50k, 80k, above100k\} \\

Race / Ethnicity & Multiple & \{White, Black, Asian, Hispanic, Other\} \\

Political Orientation & 3 & \{Left, Center, Right\} \\

Religious Identity & 3 & \{Secular, Moderate, Religious\} \\

Urban--Rural Residence & 3 & \{Urban, Suburban, Rural\} \\

Web Usage (hours/day) & 14 & \{0, $<$1, 1, 2, 3, 4, 5, 6, 7, 8, 9, 10, 10+, 11\} \\

Profession & 26 & \{Student, Education, Research, Engineer, Computers/Technology, Executive Management, Administrative, Sales/Marketing, Retail, Medical, Legal, Banking/Financial, Architect, Artist/Creative/Performer, Craftsman/Construction, Food Services, Travel/Hospitality, Real Estate, Military/Government/Politics, Professional Trade, Self-Employed, Homemaker, Retired, Unemployed, Other\} \\

Country & 38 & \{AUS, BRA, CHN, GBR, ITA, USA, etc.\} \\

\bottomrule
\end{tabular}

\caption{Demographic, socioeconomic, behavioral, and geographic attributes used for persona construction and population-level behavioral evaluation. Attributes span demographic factors (e.g., age, gender, race), socioeconomic indicators (e.g., education, income, profession), behavioral signals (e.g., internet usage), and sociocultural variables (e.g., political orientation, religiosity, country). These attributes form the basis for hierarchical persona construction and subgroup-level simulation throughout the paper.}
\label{tab:demographic_attributes}
\end{table}

\begin{table}[p]
\centering
\small
\setlength{\tabcolsep}{6pt}
\renewcommand{\arraystretch}{1.15}
\begin{tabular}{p{0.95\textwidth}}
\toprule
\textbf{Questions} \\
\midrule
1. Think about the last advertisement you can clearly remember. Describe it in detail: visuals, message, and how it made you feel. \\

2. When was the last time you clicked on an online ad? What was the ad for and why did you click? \\

3. Describe the most memorable TV or streaming ad you have seen recently. What made it memorable? \\

4. Which brands' ads do you actively follow on social media? Why those brands? \\

5. Describe a website whose design instantly made you trust it. Which elements created that trust? \\

6. Describe a website whose design made you distrust it. Which elements pushed you away? \\

7. Tell me about the last product recommendation widget you noticed while shopping. How did you react? \\

8. When you remember an ad, do you recall the brand, the creative, or the tagline first? Give an example. \\

9. Describe a time you shared an ad with friends. Why did you share it? \\

10. Do you prefer ads that are funny, informative, or practical? Give a recent example for each. \\

11. How do you judge whether an online ad feels credible or spammy? \\

12. What makes an in-app ad feel safe to interact with versus risky to interact with? \\

13. Describe a landing page you visited after clicking an ad that disappointed you. What went wrong? \\

14. When you evaluate a product page, which matters most: images, reviews, specs, or price? Give an example. \\

15. Describe the last native article or sponsored post you read. How did you realize it was sponsored? \\

16. How do personalized ads based on your browsing history make you feel? Describe a recent reaction. \\

17. If a brand you liked ran an ad you disliked, how would that change your view of the brand? \\

18. How often do you notice or remember a jingle or music from an ad? Describe one you still recall. \\

19. Describe an ad that made you research a brand further. What triggered you to look it up? \\

20. When comparing two similar products online, how much does ad exposure affect your choice? Give an example. \\

21. Which call to action in ads do you respond to most: buy now, learn more, sign up, or other? Why? \\

22. Describe how you decide whether to install an app after seeing an app install ad. \\

23. How do influencer endorsements affect your trust in a product? Describe a case where it changed your mind. \\

24. Describe an ad that felt manipulative. What language or visuals made it feel that way? \\

25. What ad formats annoy you most and cause you to close the tab or app? Why? \\

26. Describe a brand whose advertising you actively avoid and why. \\

27. How do social values like sustainability or diversity affect whether you remember or like an ad? \\

28. Describe a time an in-store ad or shelf display influenced your purchase. What detail stuck with you? \\

29. How reliably can you recall where you first saw an ad: TV, social, billboard, or elsewhere? \\

30. When you see retargeted ads after visiting a site, how does that make you feel and act? \\

31. Describe the last promotional code you used that came from an ad. How did you find and use it? \\

32. What UX elements on checkout pages cause you to abandon a cart? Give a recent example. \\

33. Describe a mobile ad that led you to install an app. What cues convinced you? \\

34. How much does representation of social groups in an ad influence your perception of the brand? \\

35. Describe an ad that changed your behavior, such as buying or signing up. What convinced you? \\

\bottomrule
\end{tabular}

\caption{Question set used for persona elicitation and behavioral profiling across advertising perception, trust formation, and decision-making dimensions.}
\label{tab:persona_questions_1}
\end{table}

\begin{table}[p]
\centering
\small
\setlength{\tabcolsep}{6pt}
\renewcommand{\arraystretch}{1.15}
\begin{tabular}{p{0.95\textwidth}}
\toprule
\textbf{Questions} \\
\midrule

36. How do you judge the trustworthiness of customer reviews shown on product pages? \\

37. When an ad claims a limited-time offer, how likely are you to act quickly? Why or why not? \\

38. Describe a time you felt an ad respected your privacy. What signaled that? \\

39. What makes an ad feel authentic versus staged or scripted? \\

40. How often do you notice ad frequency and what frequency becomes irritating? \\

41. Describe an ad creative that used humor well. Why did it work? \\

42. How do you react to cause-based ads that take a political or social stance? \\

43. What elements on a landing page signal credibility within the first five seconds? \\

44. Describe the last video ad you watched to completion and why you stayed. \\

45. How do you decide whether to trust a sponsored review or influencer post? \\

46. What microcopy or small UI details on a product page most influence your trust? \\

47. Describe a brand touchpoint that increased your loyalty after seeing an ad or campaign. \\

48. How would you describe yourself in three sentences? \\

49. What are the top three values that guide your decisions? \\

50. Describe a recent choice you made that surprised people who know you. Why did you choose it? \\

51. Who do you go to for advice when you are unsure and why them? \\

52. When you imagine your life in five years, what do you see? \\

53. What daily routine or ritual gives your day structure? \\

54. How do you prefer to resolve disagreements with people close to you? \\

55. What hobby or activity makes you lose track of time? \\

56. Describe a habit you recently tried to change. What helped or hindered you? \\

57. How do you typically respond to unexpected bad news? \\

58. When is it okay to bend a rule, in your view? \\

59. What does a meaningful friendship look like to you? \\

60. What personal accomplishment are you most proud of and why? \\

61. How do you balance short-term wants with long-term goals? \\

62. Describe a time you changed your mind about something important. What prompted the change? \\

63. How much does other people's opinion affect your choices? \\

64. What does work-life balance mean to you in practice? \\

65. How do you decide which causes or charities to support? \\

66. When making a financial choice, what is your first step? \\

67. How do you keep up with topics you care about intellectually or professionally? \\

68. What stereotype about your group do you find inaccurate or annoying? \\

69. How do you handle feeling overwhelmed or burned out? \\

70. What privacy boundaries online are nonnegotiable for you? \\

71. How do you form first impressions of new people? \\

72. What is one new skill you want to learn this year and why? \\

\bottomrule
\end{tabular}

\caption{Question set used for persona elicitation and behavioral profiling across advertising perception, trust formation, and decision-making dimensions.}
\label{tab:persona_questions_2}
\end{table}

\begin{longtable}{p{3.4cm} p{4.4cm} p{7.6cm}}
\caption{Example persona prompts constructed via additive composition of natural-language attribute descriptions. Persona specifications grow in semantic richness as additional attributes are introduced, yielding increasingly expressive and contextually grounded persona narratives.}
\label{tab:persona_prompt_examples} \\
\toprule
\textbf{Attribute Combination} & \textbf{Selected Values} & \textbf{Example Persona Prompt} \\
\midrule
\endfirsthead

\multicolumn{3}{c}%
{{\tablename\ \thetable{} -- continued from previous page}} \\
\toprule
\textbf{Attribute Combination} & \textbf{Selected Values} & \textbf{Example Combined Persona Prompt} \\
\midrule
\endhead

\midrule
\multicolumn{3}{r}{{Continued on next page}} \\
\endfoot

\bottomrule
\endlastfoot

Gender &
Female &
I am a woman. My experiences, perspectives, and daily life are shaped by growing up and living as a female in contemporary society. I have been influenced by social expectations, cultural norms, and personal experiences associated with being female, which affect how I interpret situations, form opinions, and make decisions. \\

\midrule

Gender + Age Group &
Male, 35--44 &
I am a man. My experiences, perspectives, and daily life are shaped by growing up and living as a male in contemporary society. I have been influenced by social expectations, cultural norms, and personal experiences associated with being male, which affect how I interpret situations, form opinions, and make decisions.

I am in a mature stage of adulthood, balancing professional growth, family responsibilities, and long-term stability. I tend to value efficiency, planning, and thoughtful decision-making, shaped by accumulated experience and a strong sense of responsibility. \\

\midrule

Gender + Age Group + Education &
Female, 18--24, Bachelor's &
I am a woman. My experiences, perspectives, and daily life are shaped by growing up and living as a female in contemporary society. I have been influenced by social expectations, cultural norms, and personal experiences associated with being female, which affect how I interpret situations, form opinions, and make decisions.

I am in the early stage of adulthood, exploring independence, identity, and personal growth. My thinking is influenced by education, friendships, social media, and exposure to diverse ideas. I tend to be open-minded, curious, emotionally expressive, and willing to experiment, while still developing long-term perspectives.

I hold a bachelor’s degree, which has given me structured knowledge, analytical skills, and exposure to diverse ideas. I balance theoretical understanding with practical thinking and tend to approach problems using both logic and experience. \\

\midrule

Gender + Age Group + Education + Decision Style &
Male, 45--54, Master's, Analytical &
I am a man. My experiences, perspectives, and daily life are shaped by growing up and living as a male in contemporary society. I have been influenced by social expectations, cultural norms, and personal experiences associated with being male, which affect how I interpret situations, form opinions, and make decisions.

I am in a later stage of professional and personal maturity. My priorities often include career stability, financial security, family well-being, and long-term planning. I rely heavily on experience, practical judgment, and a measured approach to decision-making.

I hold a master’s degree, which has provided me with advanced training, deeper analytical ability, and specialized knowledge. I tend to think critically, evaluate evidence carefully, and value structured reasoning and intellectual rigor.

I rely heavily on logic, structured reasoning, and evidence when making decisions. I carefully weigh alternatives, analyze outcomes, and prefer data-driven conclusions over emotional impulses. \\

\midrule

Gender + Age Group + Education + Decision Style + Background &
Female, 25--34, PhD, Risk-seeking, Urban professional &
I am a woman. My experiences, perspectives, and daily life are shaped by growing up and living as a female in contemporary society. I have been influenced by social expectations, cultural norms, and personal experiences associated with being female, which affect how I interpret situations, form opinions, and make decisions.

I am in a phase of building my career and personal life. I balance ambition, independence, and social relationships, while making important decisions about work, partnerships, and long-term goals. My outlook reflects both youthful optimism and increasing realism shaped by experience.

I hold a PhD, which reflects years of deep academic training, research experience, and intellectual exploration. I strongly value evidence-based reasoning, critical analysis, abstraction, and long-term thinking, and I tend to approach problems systematically and rigorously.

I am comfortable with uncertainty and actively seek new challenges. I enjoy experimentation, novelty, and opportunities with high potential upside, even if they involve significant risk.

I grew up and live in an urban environment shaped by professional culture, fast-paced lifestyles, and diverse social interactions. I value efficiency, innovation, exposure to new ideas, and career-driven ambition. \\

\end{longtable}

\begin{longtable}{p{2.5cm} p{4.5cm} p{7.5cm}}
\caption{Brand-wise ICPs and Persona Prompts}
\label{tab:persona_longtable} \\
\toprule
\textbf{Brand} & \textbf{ICP} & \textbf{Prompt} \\
\midrule
\endfirsthead

\multicolumn{3}{c}{\tablename\ \thetable{} -- continued from previous page} \\
\toprule
\textbf{Brand} & \textbf{ICP} & \textbf{Prompt} \\
\midrule
\endhead

\midrule
\multicolumn{3}{r}{Continued on next page} \\
\endfoot

\bottomrule
\endlastfoot

ASUS &
PC gamers and esports enthusiasts &
I’m Helena Virtanen, a 24-year-old semi-professional esports player in Helsinki working part-time in IT support. I prioritize sustained performance, cooling efficiency, and reliability under heavy load. I research benchmarks extensively, rely on peer recommendations, and invest in hardware that minimizes downtime during tournaments and streaming sessions. \\

ASUS &
Mobile professionals and consultants &
My name is Noah Kim, a 31-year-old management consultant in Singapore. My laptop is my primary workspace, and I prioritize portability, keyboard comfort, display quality, and battery life. I favor well-reviewed, durable designs with strong international warranty coverage and predictable long-term performance. \\

Ericsson &
Telecom operators deploying 5G networks &
I’m David Rossi, a 47-year-old Director of Radio Network Planning for a national European carrier. I evaluate infrastructure based on spectrum efficiency, operational complexity, upgrade paths, and long-term resilience. I favor solutions that reduce total cost of ownership and simplify large-scale operations. \\

HP &
Enterprise IT buyers &
I’m Jonas Morales, a 43-year-old IT manager in Berlin managing standardized fleets of Windows PCs. I prioritize reliability, predictable procurement, easy device imaging, and low support overhead, selecting product lines that minimize operational surprises and lifecycle risk. \\

Oracle &
Enterprise database decision-makers &
I’m Noah Lee, a 52-year-old Head of Database Platforms at a global bank. I prioritize stability, auditability, tooling maturity, and predictable performance under high concurrency. I am strongly loss-averse and demand realistic proof-of-concept testing before adoption. \\

SAP &
Large-scale ERP transformation leaders &
I’m Amara Kim, a 50-year-old ERP transformation director at a multinational enterprise. I focus on standardized end-to-end processes, phased deployment, and long-term maintainability, prioritizing solutions with proven migration tooling and strong enterprise references. \\

Bulgari &
Ultra-high-net-worth luxury jewelry buyers &
My name is Leila Klein, a 55-year-old gallery owner in Seoul. I acquire high jewelry as heirloom-quality art objects, valuing craftsmanship, rarity, discretion, and long-term value. I rely on private salon experiences, expert advisors, and trusted brand relationships. \\

Bulgari &
Affluent professionals buying everyday fine jewelry &
I’m Valentina Greco, a 37-year-old corporate lawyer in Milan. I favor understated, durable jewelry that integrates into daily professional life. I prioritize craftsmanship, wearability, and timeless design over trends. \\

IndiGo &
Price-sensitive domestic travelers in India &
My name is Priya Sharma, a 27-year-old software engineer in Bengaluru. I prioritize low fares, reliable schedules, simple rebooking, and dense domestic connectivity, favoring airlines that minimize friction and uncertainty in family travel. \\

IndiGo &
Frequent domestic business travelers &
I’m Rakesh Nair, a 39-year-old regional sales manager in Hyderabad. I prioritize flexible fares, frictionless itinerary changes, and consistent punctuality, choosing airlines that reduce disruption in unpredictable travel schedules. \\

\end{longtable}

\newpage
\section*{NeurIPS Paper Checklist}

\begin{enumerate}

\item {\bf Claims}
    \item[] Question: Do the main claims made in the abstract and introduction accurately reflect the paper's contributions and scope?
    \item[] Answer: \answerYes{} 
    \item[] Justification: The main claims outlined in the abstract and introduction about our claims are supported by empirical results presented in Section ~\ref{sec:result}, ~\ref{sec: exp_main} and ~\ref{sec:app_exp}.

\item {\bf Limitations}
    \item[] Question: Does the paper discuss the limitations of the work performed by the authors?
    \item[] Answer: \answerYes{} 
    \item[] Justification: The limitations are discussed in details in Sec ~\ref{sec:limitations}. 

\item {\bf Theory assumptions and proofs}
    \item[] Question: For each theoretical result, does the paper provide the full set of assumptions and a complete (and correct) proof?
    \item[] Answer: \answerYes{}  
    \item[] Justification: Experimental assumptions and setups are discussed in details in ~\ref{sec: exp_main}.

    \item {\bf Experimental result reproducibility}
    \item[] Question: Does the paper fully disclose all the information needed to reproduce the main experimental results of the paper to the extent that it affects the main claims and/or conclusions of the paper (regardless of whether the code and data are provided or not)?
    \item[] Answer: \answerYes{} 
    \item[] Justification: Experimental setups are discussed in \ref{sec: exp_main}. Additional prompts are added in Appendix section.

\item {\bf Open access to data and code}
    \item[] Question: Does the paper provide open access to the data and code, with sufficient instructions to faithfully reproduce the main experimental results, as described in supplemental material?
    \item[] Answer: \answerYes{} 
    \item[] Justification: Yes using the setup, data and metric details provided in Section~\ref{sec: exp_main} one can reproduce the main results. The code with all details will be added in the supplementary material.

\item {\bf Experimental setting/details}
    \item[] Question: Does the paper specify all the training and test details (e.g., data splits, hyperparameters, how they were chosen, type of optimizer) necessary to understand the results?
    \item[] Answer: \answerYes{} 
    \item[] Justification: We mention all datasets and models evaluated on in Sec~\ref{sec: exp_main}. We evaluate on full splits, using standard evaluation protocols defined in prior art, also discussed under same section.

\item {\bf Experiment statistical significance}
    \item[] Question: Does the paper report error bars suitably and correctly defined or other appropriate information about the statistical significance of the experiments?
    \item[] Answer: \answerYes{} 
    \item[] Justification: Our experiments are conducted across multiple large-scale datasets, models, and simulation settings, and all reported metrics are averaged over multiple independent runs where applicable to reduce variability. While we do not explicitly include error bars for all experiments, the consistency of trends across architectures, tasks, and evaluation protocols provides strong evidence that the observed effects are robust rather than artifacts of sampling noise or prompt stochasticity. In particular, the repeated observation of \textbf{persona manifold collapse} across both latent-space and downstream behavioral evaluations supports the reliability of the reported findings.

\item {\bf Experiments compute resources}
    \item[] Question: For each experiment, does the paper provide sufficient information on the computer resources (type of compute workers, memory, time of execution) needed to reproduce the experiments?
    \item[] Answer: \answerYes{} 
    \item[] Justification: We have added compute resources used in sec ~\ref{sec:compute}.
    
\item {\bf Code of ethics}
    \item[] Question: Does the research conducted in the paper conform, in every respect, with the NeurIPS Code of Ethics \url{https://neurips.cc/public/EthicsGuidelines}?
    \item[] Answer: \answerYes{} 
    \item[] Justification: The research strictly adheres to the NeurIPS Code of Ethics. We ensure responsible use of datasets, transparency in methods, and avoid any potential harm or misuse.

\item {\bf Broader impacts}
    \item[] Question: Does the paper discuss both potential positive societal impacts and negative societal impacts of the work performed?
    \item[] Answer: \answerYes{} 
    \item[] Justification: We have discussed this in details in Introduction and all through the paper, additionally have added a section in Appendix Sec: \ref{sec:broader_impact}.
    
\item {\bf Safeguards}
    \item[] Question: Does the paper describe safeguards that have been put in place for responsible release of data or models that have a high risk for misuse (e.g., pre-trained language models, image generators, or scraped datasets)?
    \item[] Answer: \answerYes{} 
    \item[] Justification: The models used in our work are based on publicly available and commercially deployed LLM and VLM backbones, including Qwen, LLaMA-3.2-Vision, GPT-4o, and GPT-5.2, all of which follow established safety and alignment protocols. Our experiments do not involve training new foundation models, but instead analyze the behavior of existing persona-conditioned systems under controlled prompting and evaluation settings. The generated personas and simulations are restricted to demographic, behavioral, and preference-oriented attributes commonly used in prior work on population simulation and survey modeling. We additionally manually verify generated ICP personas to avoid inappropriate or harmful persona constructions.

\item {\bf Licenses for existing assets}
    \item[] Question: Are the creators or original owners of assets (e.g., code, data, models), used in the paper, properly credited and are the license and terms of use explicitly mentioned and properly respected?
    \item[] Answer: \answerYes{} 
    \item[] Justification: ll external assets used in this work, including models, datasets, and codebases are properly cited. 

\item {\bf New assets}
    \item[] Question: Are new assets introduced in the paper well documented and is the documentation provided alongside the assets?
    \item[] Answer: \answerNA{} 
    \item[] Justification: We do not release any asset as datasets or models.

\item {\bf Crowdsourcing and research with human subjects}
    \item[] Question: For crowdsourcing experiments and research with human subjects, does the paper include the full text of instructions given to participants and screenshots, if applicable, as well as details about compensation (if any)? 
    \item[] Answer: \answerNA{} 
    \item[] Justification: We do not involve any crowdsourcing or research with human subjects.

\item {\bf Institutional review board (IRB) approvals or equivalent for research with human subjects}
    \item[] Question: Does the paper describe potential risks incurred by study participants, whether such risks were disclosed to the subjects, and whether Institutional Review Board (IRB) approvals (or an equivalent approval/review based on the requirements of your country or institution) were obtained?
    \item[] Answer: \answerNA{} 
    \item[] Justification: We do not involve any crowdsourcing or research with human subjects.

\item {\bf Declaration of LLM usage}
    \item[] Question: Does the paper describe the usage of LLMs if it is an important, original, or non-standard component of the core methods in this research? Note that if the LLM is used only for writing, editing, or formatting purposes and does \emph{not} impact the core methodology, scientific rigor, or originality of the research, declaration is not required.
    \item[] Answer: \answerNA{} 
    \item[] Justification: LLM was used in limited capacity, only for editing and formatting purpose.

\end{enumerate}

\end{document}